\documentclass[10pt]{article}

\usepackage{amsmath}
\usepackage{amssymb}
\usepackage{dsfont}

\usepackage{graphicx}

\usepackage{cite}

\usepackage{color}

\usepackage[labelfont=bf,labelsep=period,justification=raggedright]{caption}

\makeatletter
\renewcommand{\@biblabel}[1]{\quad#1.}
\makeatother

\date{}

\begin{document}

\begin{flushleft}
{\Large
\textbf{Reconstructing fiber networks from confocal image stacks\\}
}
Patrick Krauss, 
Claus Metzner$\ast$, 
Janina Lange, 
Nadine Lang, 
Ben Fabry
\\
Dept. of Physics, Biophysics Group, Friedrich-Alexander University, Erlangen, Germany
\\
$\ast$ E-mail: claus.metzner@gmx.net
\end{flushleft}

\section*{Abstract}

We present a numerically efficient method to reconstruct a disordered network of thin biopolymers, such as collagen gels, from three-dimensional (3D) image stacks recorded with a confocal microscope. Our method is based on a template matching algorithm that simultaneously performs a binarization and skeletonization of the network. The size and intensity pattern of the template is automatically adapted to the input data so that the method is scale invariant and generic. Furthermore, the template matching threshold is iteratively optimized to ensure that the final skeletonized network obeys a universal property of voxelized random line networks, namely, solid-phase voxels have most likely three solid-phase neighbors in a $3\times3$ neighborhood. This optimization criterion makes our method free of user-defined parameters and the output exceptionally robust against imaging noise.



\section*{Introduction}

\begin{figure}[!htb]
\centering
\includegraphics[width=0.8\linewidth]{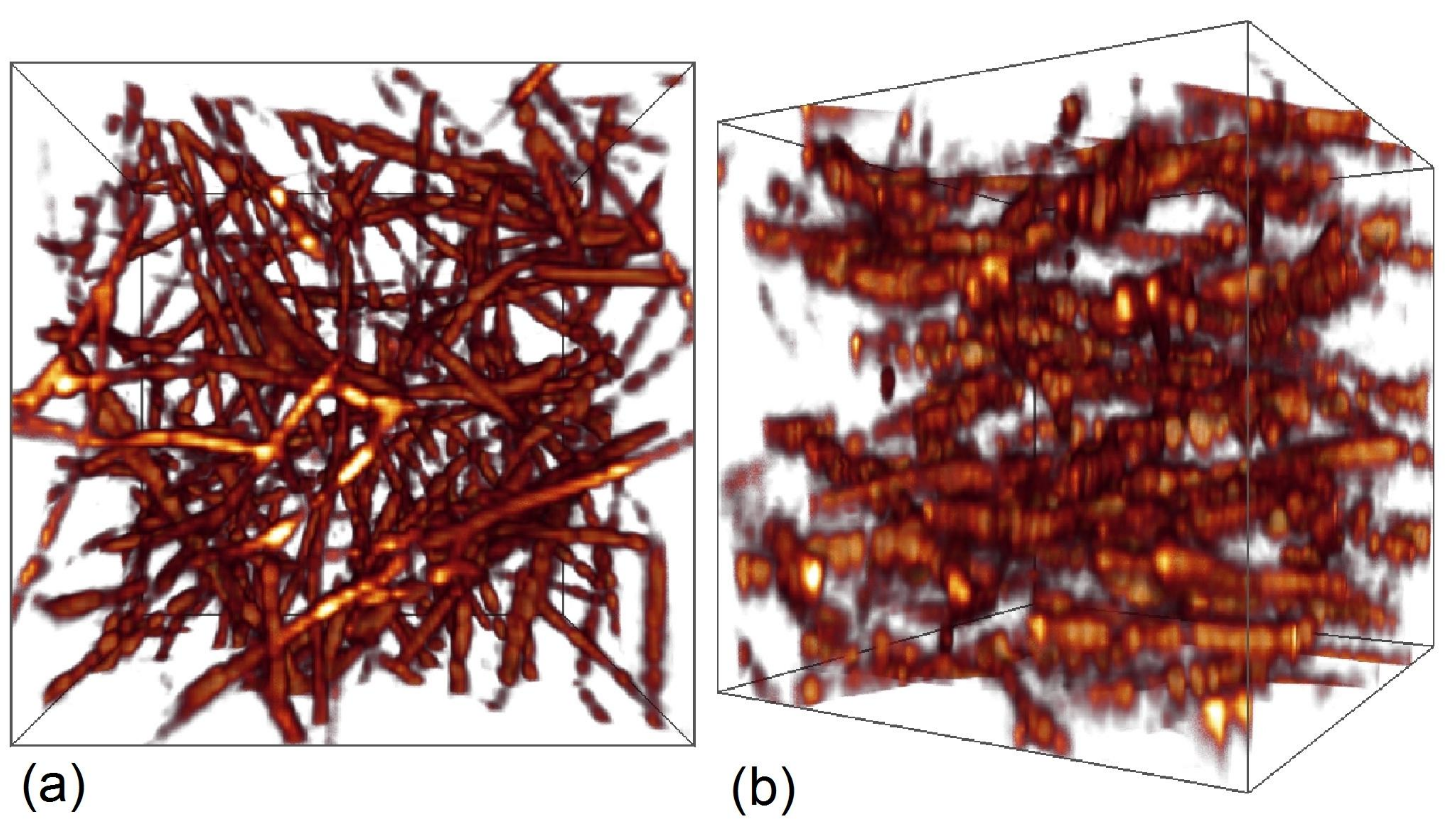}
\caption{\label{fig:network} \textbf{A cube of collagen gel.} Dimensions $(32 \times 32 \times 34) \mu m^3$, concentration 1.2\,$\frac{\rm{mg}}{\rm{ml}}$, recorded with confocal reflection microscopy and without any image processing. (a) Top view, i.e. in z-direction. (b) Side view. The lateral (x-, y-direction) resolution of the fibers is considerably better than the vertical (z-direction) resolution, due to the anisotropic point spread function. In addition, only fiber segments that run in small angles to the imaging plane are visible, the so-called blind spot effect.}
\end{figure}

Many biological materials, such as the cytoskeleton or the extracellular matrix, self-organize into complex networks by the polymerization of protein molecules into fibrils (Fig.\,\ref{fig:network}). If the thickness of the fibrils is negligible compared to the pore size, the resulting structure can be mathematically described as a disordered line network. In general, the functional properties of these networks, such as their mechanical stiffness on the macroscopic scale, or their permeability for diffusing particles and for actively migrating cells on a microscopic scale, depend on the geometrical details of the microscopic network structure. In order to study the relationship between structure and function, it is therefore important to extract, or reconstruct, the 3D network structure from image stacks. One aspect of the reconstruction is the binarization of the intensity values of the image stack, so that each voxel is assigned one of two possible values, corresponding either to the solid phase (1, collagen fibers) or the liquid phase (0, surrounding medium). Another aspect of the reconstruction is the skeletonization, so that the optically broadened fibers are reduced to their central (medial) axis, with a width of only one voxel.

While most of the standard reconstruction methods realize the two aspects of reconstruction (i.e. binarization and skeletonization) in a two-step process, our template matching method achieves both aspects in a single step. This new method avoids the problem of chosing an arbitrary intensity threshold for the binarization. Instead, the template matching algorithm automatically adapts to the input data such that within the reconstructed fraction of solid-phase voxels the most probable number of next neighbors equals three. This represents a universal property of voxelized line networks.

\paragraph{Criteria for reconstruction methods}

We regard it essential to define the following criteria for our reconstruction method: (1) The method needs to be free of user-adjustable parameters, and (2) be insensitive to variations in the input data quality. To test this criterion we image collagen networks under a wide range of different confocal microscope settings such as amplifyer gain and laser outlet power. The method (3) must be able to correctly reconstruct known networks. To simulate realistic conditions, these networks are convoluted with a point spread function of the imaging system, and different levels of noise are added. 

\paragraph{Existing reconstruction methods}

A vast variety of methods can be used for the reconstruction problem. A large class of these methods works with two separate steps of binarization and skeletonization \cite{pud98, ma96, lee94, pro07, ren09, wan07}. The simplest way to binarize an image stack is by comparing each individual voxel's grayscale with a threshold value $\theta$ and to assign all voxels that are brighter than $\theta$ to the solid phase. As we shall demonstrate below, this method naturally leads to binarized arrays with many artifacts, i.e. false positive and false negative voxels. This includes the simple removal of isolated solid-phase voxels, which can result from noise or dirt particles in the medium. More demanding, it requires the thinning of the broadened binarized fibers to their medial axis of one voxel diameter. Binarization can also lead to the disintegration of fibers, so that closing methods, consisting of dilatation with subsequent erosion steps \cite{soi99, sze10}, have to be applied as well. Another class of methods is based on edge detection with convolution kernels, using, for example, Laplace filters or Sobel operators \cite{soi99, sze10}. This class of methods is also plagued with the production of artifacts that have to be removed afterwards. Finally, there are the class of learning algorithms, such as vector clustering methods \cite{gan07} and neural networks \cite{bis96}, for example the k-means algorithm \cite{bra00} or RBF networks \cite{how07}. A significant advantage of such methods is their ability to automatically adapt to the specific properties of the data at hand. Our proposed template matching algorithm generates its template patterns automatically from the data and can therefore be considered as a learning algorithm as well.

\paragraph{Problems with existing methods}

A detailed summary and comparison of all reconstruction methods is beyond the scope of this paper. Instead, we shall briefly consider the simple example of global threshold binarization and discuss some of its fundamental shortcomings. This will be useful to highlight the advantages of the template matching method proposed later.

We start with an image stack recorded by reflection microscopy. Let us assume that the grayscales of the image stack are coded with 8 bits, i.e. all brightness values $B$ are in the range $B \in [0,255]$, with $B=0$ corresponding to completely dark (black) and $B=255$ to maximum bright (white) voxels. In our setup (Leica SP5X confocal microscope), a typical distribution $p(B)$ of brightness values has a sharp peak around $B=15\pm5$ and a very flat tail towards large values (Fig.\,\ref{fig:SURROGATE}a). The reasonable range of binarization thresholds $\theta$ is located somewhere within this tail. However, the distribution $p(B)$ itself offers no hint as to where the optimum threshold point should be set.

To characterize different network reconstruction methods, we use artificially generated image stacks. This requires realistic models of both, the line network itself and its transformation into cross-sectional images by the microscope. As described in more detail in the Methods section, we use a "Mikado" model for the line network, where straight lines of fixed lengths and isotropic orientations are distributed throughout the volume with a homogeneous density \cite{met11}. To model the imaging process, we take into account the broadening (simulated by a convolution with a point spread function), the blind spot effect (i.e. a gradual darkening of steep fibers) \cite{jaw09} and the addition of random noise. The resulting image stacks have statistical properties almost indistinguishable from measured image stacks (Fig.\,\ref{fig:SURROGATE}), but with the advantage that the underlying mathematical line network is known precisely.

Is it possible to perfectly reconstruct the original line network using global threshold binarization? This would require the existence of a threshold $\theta$, such that all fluid-phase voxels have brightnesses below and all solid-phase-voxels brightnesses above this threshold. However, when we use our synthetic stack and plot the brightness distributions $p_S(B)$ and $p_F(B)$ of the two phases separately, we find in general two peaks with a significant overlap (Fig.\,\ref{fig:brightnesshistoseperate}). This means that no global threshold can be found, even in principle, for separating the two phases, without also producing some false positive and false negative voxels.

The voxels with brightnesses in the overlap interval include, for example, isolated bright points due to noise. It would be relatively easy to remove such cases in a subsequent post-processing step. More problematic is that the overlap interval also includes liquid phase voxels from the narrow gaps between two fibers, which have been raised in brightness beyond the threshold by the superposition of the fibers' point spread functions. This effect would lead to a merging of the two close-by fibers in the binarized image and would require a much more sophisticated procedure to be repaired. Finally, the overlap region includes voxels of fiber segments that are more vertically oriented, and therefore too dark, to exceed the threshold, because of the blind spot problem \cite{jaw09}. We note that a human observer could still recognize such dark fiber segments quite easily.

Taken together, the threshold binarization has some fundamental limitations. To a certain extent, the method can be improved by using variable thresholds, which take into account the local brightness conditions in the environment of each voxel to be binarized. This, however, can already be viewed as a first step towards a template matching method that will be discussed in the following.

\paragraph{Template matching in line networks}

Template matching methods recognize specific image parts within larger image stacks by comparing features, e.g. the brightness patterns, of small sub volumes of the stack with the brightness pattern of pre-defined templates. The templates incorporate the a-priori-knowledge about the features to be found. In the case of line networks, the templates would contain short line segments, that are oriented in arbitrary directions.

The number of required templates turns out to be impractically large in 3D. However, the situation is much simpler when only 2D cross-sections are used for the template matching: The vertical cross-section of a broadened line segment with a plane is a elliptical spot of finite size that can be easily recognized by 2D template matching (Fig.\,\ref{fig:SLICES}a). The shape of the spot will vary slightly as the angle of intersection becomes less than 90 degrees. For angles less than 45 degrees, the distortion of the spot can become too large to match the template (Fig.\,\ref{fig:SLICES}b), but in this case the same line segment can be easily recognized by its intersection with a perpendicular plane. Therefore, all line segments (solid voxels) can be detected by sequentially scanning through the x-, y- and z-planes of the sample volume. As shown below this binarization method turns out to be much more reliable and robust than the simple threshold method.

The similarity between the cross-sectional brightness pattern and the template will be largest if the template is located exactly at the center of the finite spot. Therefore, the medial axes of the broadened line segments can be identified as local minima of the mismatching measure. In this way, the 2D template matching simultaneously achieves a skeletonization of the broadened fibers.

We note that this method meets the design criteria imposed before. In order to eliminate all internal parameters (1), we have implemented an automatic template generator, which is entirely based on the input image stack and requires no user intervention. We will demonstrate in the Results section that our method is also robust with respect to the quality of the input data (2) and yields reconstructions that reproduce simulated line networks almost perfectly (3). The algorithm is implemented in C++ to achieve fast execution times (about 12 minutes on a standard notebook for reconstructing a $(512\times512\times597)$-voxel stack) and is available as a supplement to this publication.

\section*{Methods}


\paragraph{Generation of surrogate data sets}

Since the true shape of the underlying fiber network of real grayscale data sets is unknown, we artificially generated surrogate data sets to validate the performance of our algorithm. Firstly we created idealized line networks using a "Mikado" model. Straight lines of fixed lengths and isotropic orientations are distributed throughout the volume with a homogeneous density \cite{met11}. Binary surrogate data sets are derived from these parameterized networks by a voxelization operation. Then we simulate the imaging process to transform these binary data sets into grayscale data sets.

Initially $N$ sets of parameters representing $N$ lines are randomly generated. Each parameter set contains uniformly distributed Cartesian coordinates of a line's center point $(x_c,y_c,z_c)$, uniformly distributed azimuthal angle $\varphi \in \left[-\pi,\pi\right]$ with $p(\varphi)=\frac{1}{2\pi}$ and polar angle $\vartheta \in \left[0,\pi\right]$ with $p(\vartheta)=\frac{\sin\vartheta}{2}$.
This is an efficient way of representing a line network, since for each arbitrary point within the volume one can unambiguously determine whether or not the point is located on any line.

To derive a binary data set from the parameterized network the whole volume is divided into distinct volume elements representing the voxels. Initially all voxels within the 3D-array are set to value 0 (fluid phase). Starting at the first center point, the line is traced in opposite directions according to its orientation $(\varphi,\vartheta)$ until the half length of the line is reached for each direction. While tracing the line, every voxel corresponding to a touched volume element is set to value 1 (solid phase). This process is repeated for all $N$ sets of line parameters. 

To convert the binary data set into a 8-bit-grayscale data set, we apply a process called numeric blurring which simulates the imaging process. Numeric blurring is done in five subsequent steps:
\begin{enumerate}

  \item The binary 3D-array of voxels is converted into a grayscale 3D-array by setting all voxels with value 1 to brightness values $B \in [0,255]$ according to the polar angle $\vartheta$ of the line to which they belong. This procedure replicates the blind spot effect (i.e. a gradual darkening of steep fibers) as described in \cite{jaw09}.

	\item A small number of dark voxels is randomly selected and set to brightness values $>200$. This simulates dirt particles within the fluid phase that appear as isolated bright fluctuations in the original microscope images.
		
	\item The 3D-array of voxels is convoluted with an anisotropic Gaussian to simulate the point spread function: 
	\begin{align}
    B(x,y,z) &:= (B \ast psf)(x,y,z) \\
    (B \ast psf)(x,y,z) &= \iiint\limits_{}B(x',y',z')psf(x-x',y-y',z-z')dx' \, dy' \, dz'
  \end{align}
  where $B(x,y,z) \in \left[0,255\right]$ is brightness of voxel $(x,y,z)$. The point spread function is defined as \\
  $psf(x-x',y-y',z-z') = \exp\left(-\left(\frac{(x-x')^{2}}{\sigma_{x}^{2}}+\frac{(y-y')^{2}}{\sigma_{y}^{2}}+\frac{(z-z')^{2}}{\sigma_{z}^{2}}\right)\right)$ with $\sigma_x=\sigma_y<\sigma_z$ according to the characteristics of confocal microscopy.
		
	\item A Gaussian distributed random variable is added to each voxel to simulate noise\footnote{We are aware that photon shot noise would not be normally distributed. However, as shown below, the resulting statistical properties of the surrogate stacks agree almost perfectly with measured data.}. 
	
	\item Finally, all brightness values are rescaled to values from 0 to 255 by affine transformation.
	
\end{enumerate}

\begin{figure} [!htb]
\begin{center}
\includegraphics[width=1.0\linewidth]{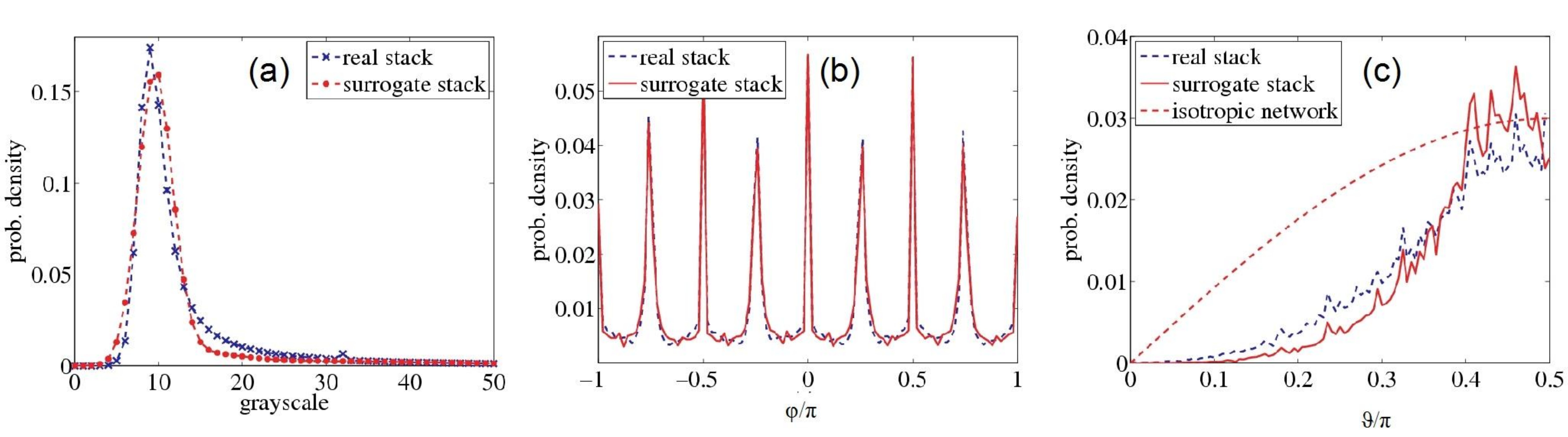}
\end{center}
\caption{\textbf{Statistical properties of real and surrogate image stacks.} (a) Comparison of the brightness distributions in the real and surrogate image stacks. Both distributions are similar. (b) and (c) show angular distributions of the fiber segments. (b) Typical distributions of azimuthal angles $\varphi$ in a real and a surrogate data set. The distributions are almost indistinguishable. The peaks are a result of voxelization. The principal directions, corresponding to the x- and y-direction, as well as the principal diagonals are over-represented in short fiber segments and lead to maxima at $\varphi=0, \pm\frac{\pi}{4}, \pm\frac{\pi}{2}, \pm\frac{3\pi}{4}, \pm\pi$. (c) Typical distributions of polar angles $\vartheta$ in a real and a surrogate data set. Again, the distributions are similar. Compared to an ideal isotropic network with $p(\vartheta)\propto \sin(\vartheta)$, polar angles smaller than $\frac{\pi}{2}$ are increasingly suppressed due to the blind spot effect of confocal reflection microscopy \cite{jaw09}.}
\label{fig:SURROGATE}
\end{figure}

We have analyzed the statistical properties of the resulting grayscale data sets\footnote{It is possible to determine the direction vector of a short fiber segment, even from its voxelized representation, by treating the brightness distribution as a mass distribution, computing and diagonalizing the moment of inertia tensor, and finding the principal component axis of mimimal inertia. This principal component corresponds to the direction of the locally straight line segment. Several conditions have to be met when analyzing a given small test volume: First, the fluid background of the fiber segment in the test volume should have a much smaller (ideally zero) brightness/mass than the fiber itself. It is therefore advisable to use already reconstructed stacks for this analysis. Second, the test volume should contain enough solid voxels to clearly define a single fiber segment. Third, the test volume should not be so large to contain several line-like objects with different directions. We have therefore used the following method: A number of PMX=$10^5$ spherical test volumes of radius r=3.0 vu (voxel unit, i.e. linear size of a single voxel) have been chosen randomly throughout the reconstructed stack, ensuring that each test volume contains at least NMIN=5 solid voxels. Inside each sphere, the voxels were treated as mass points, located at the voxel centers, and with constant mass m=1 for all solid and m=0 for all liquid voxels. After determining the easy axis of the inertia tensor, the corresponding unit direction vector was computed. Note that this vector does not depend on the exact position of the test sphere's center, as long as the same solid voxels are enclosed. From this Cartesian vector, the azimuthal angle $\varphi$ and the polar angle $\vartheta$ in spherical coordinates were computed. Finally, histograms were generated for $\varphi$ and $\vartheta$.}. They are almost indistinguishable from real data sets imaged with confocal reflection microscopy (Fig.\,\ref{fig:SURROGATE}), but with the advantage that the underlying mathematical line network is known in detail.

\begin{figure} 
\begin{center}
\includegraphics[width=0.5\linewidth]{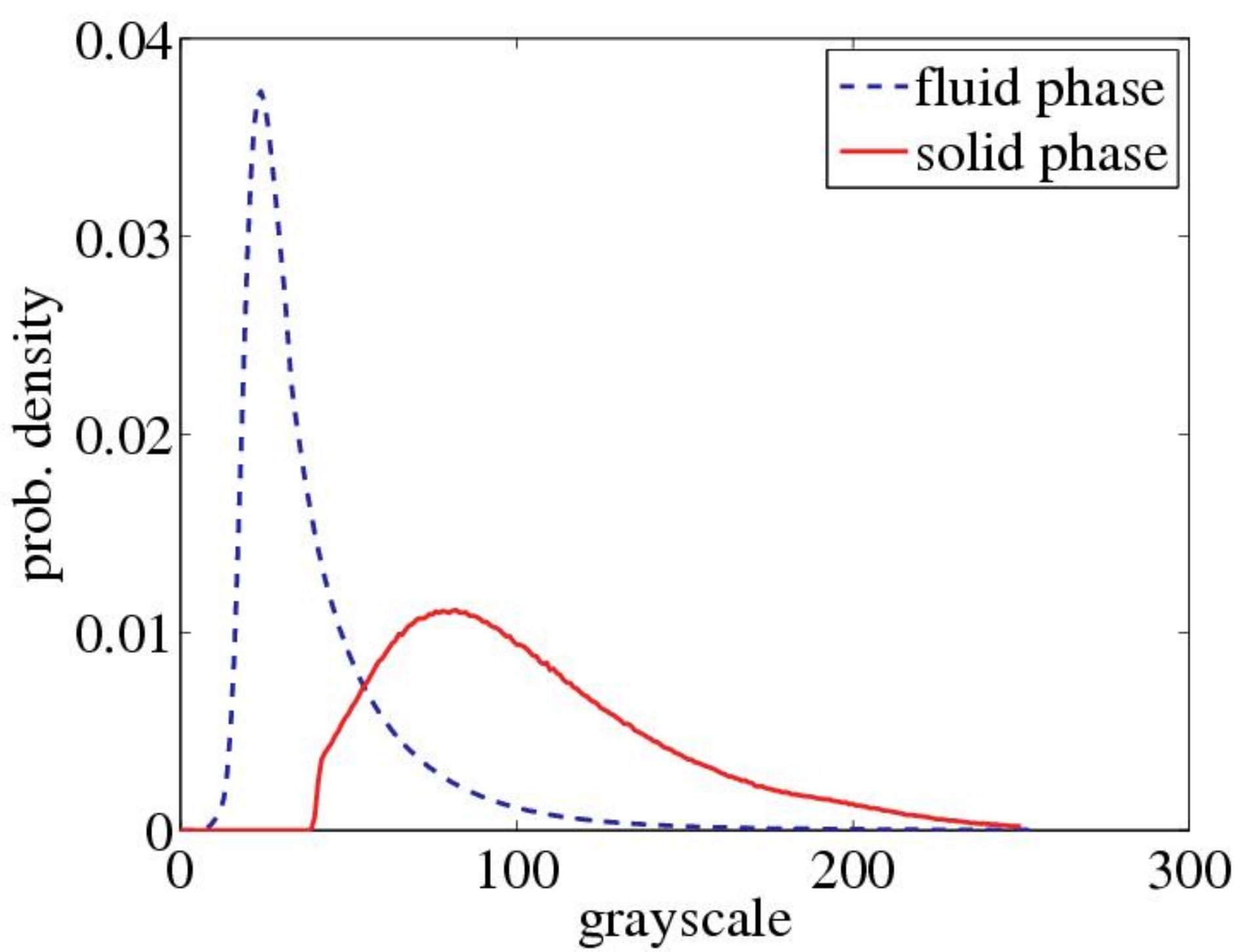}
\end{center}
\caption{\textbf{Brightness distributions $p_S(B)$ and $p_F(B)$ of the solid and the fluid phase.} The two peaks show a significant overlap. This clearly points out that no global threshold can be found, even in principle, for separating the two phases, without also producing some false positive and false negative voxels.}
\label{fig:brightnesshistoseperate}
\end{figure}

When we plot the brightness distributions $p_S(B)$ and $p_F(B)$ of the two phases (solid and fluid) separately, we find in general two peaks with a significant overlap (Fig.\,\ref{fig:brightnesshistoseperate}). This means that no global threshold can be found, even in principle, for separating the two phases, without also producing some false positive and false negative voxels.

\paragraph{Preprocessing}

Some data sets show a z-dependence of mean grayscale. The average brightness of each z-slice $\mu(z)$ is not constant in all layers, but rather decreases for deeper located slices in the stack. This effect is caused by scattering and absorption of light by collagen fibers and ambient medium.
For compensation the data set is firstly normalized to an equal global mean grayscale, disregarding statistical fluctuations:
\begin{align}
B_{xyz} &:= B_{xyz}-\mu(z) & \forall \ x,y,z \\
\intertext{where $ \mu(z)=\frac{1}{N_x N_y}\sum\limits_{i=1}^{N_x}\sum\limits_{j=1}^{N_y} B_{ijz}$}
\intertext{and secondly rescaled to values $\left[0,255\right]$ by affine transformation:}
B_{xyz} &:= 255\cdot\frac{B_{xyz} - \min\left\{B_{ijk}\right\}} {\max\left\{B_{ijk}\right\} - \min\left\{B_{ijk}\right\}} & \forall \ x,y,z,i,j,k
\end{align}

\paragraph{Automatic template generation}

\begin{figure} [!htb]
\begin{center}
\includegraphics[width=0.35\linewidth]{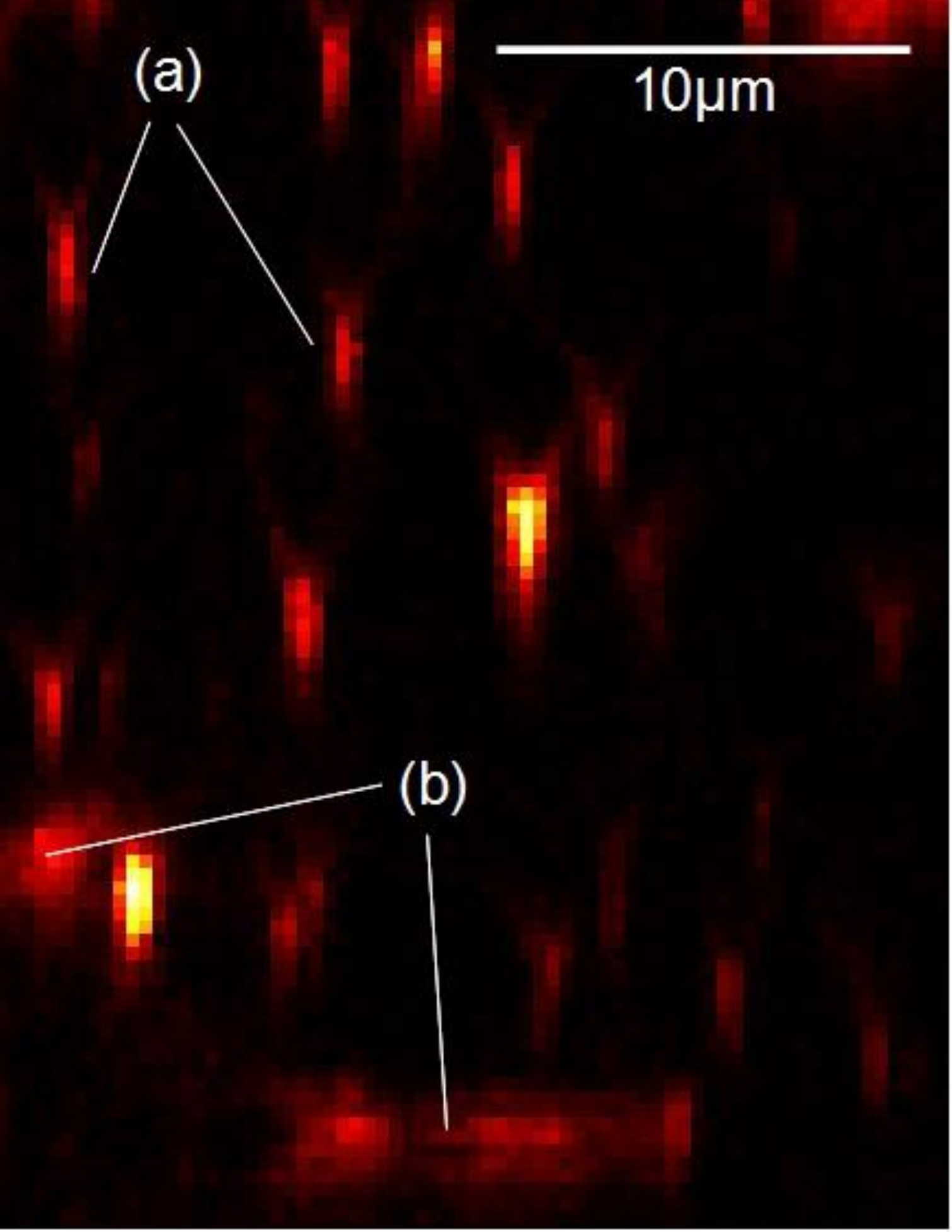}
\end{center}
\caption{\textbf{2D cross-section of a 3D image stack.} (a) The vertical cross-section of a broadened line segment with a plane is a elliptical spot of finite size that can be easily recognized by 2D template matching. (b) The shape of the spot is varying slightly as the angle of intersection becomes less than 90 degrees. For angles less than 45 degrees, the distortion of the spot can become too large to match the template.}
\label{fig:SLICES}
\end{figure}

The number of required templates turns out to be impractically large in 3D. However, the situation becomes much easyer when only 2D cross-sections are used for the template matching: The vertical cross-section of a broadened line segment with an image plane is a ellitpical spot of finite size that can be easily recognized by 2D template matching (Fig.\,\ref{fig:SLICES}).

The templates are derived automatically from the gray scale data set. The described process requires neither any knowledge of the point spread function's characteristics nor any user interactions. Since fiber detection is performed in three directions (x, y and z) and the point spread function may be anisotropic, three different templates $\mathbf T_x, \mathbf T_y, \mathbf T_z$ are required. One for each direction.
\begin{enumerate}

\item A sufficient large sample of voxels ($10^5$) is picked out randomly. In order to minimize computation time only voxels with intensities greater than the global mean intensity $\mu$ were accepted.

\item Each voxel is the center of three small 2D images ($I_x, I_y, I_z$), where the first image is located in the yz-, the second in the xz- and the third in the xy-plane. The images' height $H$ and width $W$ are initially set to arbitrary small values. The adaptive resizing process is described in a following section.

\item These small images are represented as $H\times W$ matrices $\mathbf P_1,\mathbf P_2,\ldots,\mathbf P_N$, with matrix entries corresponding to voxel brightnesses. Hence three sets of training patterns $\left\{\mathbf P^x_i\right\}, \left\{\mathbf P^y_i\right\}, \left\{\mathbf P^z_i\right\}$ are obtained to compute the three different templates.

\item From all training patterns belonging to a set, the weighted average pattern $\mathbf A$ is computed:
\begin{equation}
\mathbf{A}:=\frac{\sum\limits_i w_i\mathbf P_i}{\sum\limits_i w_{i}}
\end{equation}
where $w_i$ is the central entry of matrix $\mathbf P_i$. 

\item To become independent from absolute brightnesses the global mean gray scale $\mu$ is subtracted from each matrix entry.

\item Finally the matrices $\mathbf{A_x}, \mathbf{A_y}, \mathbf{A_z}$ are normalized to obtain the templates $\mathbf T_x, \mathbf T_y, \mathbf T_z$:
\begin{equation}
\mathbf{T}:=\frac{1}{\left\|\mathbf{A}\right\|} \, \mathbf{A} 
\end{equation}

\end{enumerate}

\paragraph{Fiber detection process}

As previously mentioned the fiber detecting process in the gray scale input data set is performed subsequently for all three directions x, y and z. During these three cycles three temporary binary output data sets $\mathbf O_x, \mathbf O_y, \mathbf O_z$ are obtained, having each voxel labeled with one of two possible values 0 (fluid phase) or 1 (collagen phase).
The similarity between a given cross-sectional brightness pattern and the corresponding template will be largest if the template is located exactly at the center of the finite spot. Therefore, the medial axes of the broadened line segments can be identified as local minima of the mismatching measure. The fiber detection process includes the following steps: 
\begin{enumerate}

\item Initially all voxels of the first x-slice (yz-plane) with brightnesses larger than the mean gray scale $\mu$ are investigated. This restriction is only made to decrease computation time. Without this restriction the fraction of additionally detected fiber voxels is less than $10^{-5}$. This small fraction is negligible since it does not significantly affect the resulting network properties on the one hand. On the other hand, performance time would be increased drastically by a factor of 5, because 80\% of all voxels are darker than the mean brightness. Consequently all voxels with gray scales smaller than $\mu$ are set to 0 in the binarized output data set $\mathbf O_x$.

\item The voxels to be investigated are taken as centers of small 2D-images located in the yz-plane. These images are represented as matrices with the same size as the corresponding template. These matrices are the unknown patterns to be compared with the template. We call them search patterns $\mathbf S_n$.

\item From each entry of $\mathbf S_n$ the local mean value $\mu_n$ is subtracted.

\item The search patterns are normalized:
\begin{equation}
\mathbf{S_n}:=\frac{1}{\left\|\mathbf{S_n}\right\|} \, \mathbf{S_n} 
\end{equation}

\item Now the mismatch $d_n$ between search patterns $\mathbf{S_{n}}$ and template pattern $\mathbf{T_{x}}$ for detection in x-direction is computed. The smaller $d_n$, the more probable the central voxel of the search pattern belongs to the collagen phase. As mismatch measuring metric we chose the Euclidean distance in feature space. Hence we call $d_n$ the matching distance.
\begin{equation}
d_{n}:=\left\|\mathbf{S_{n}}-\mathbf{T_x}\right\|=\sqrt{\sum\limits_{i}\sum\limits_{j}\left(s_{ij}^{n}-t_{ij}^{x}\right)^{2}}
\end{equation}
Since all matrices are normalized, the range of $d_n$ is $\left[0,2\right]$ where $d_n=0$ means perfect matching.

\item The calculated matching distances $d_n$ are compared with the threshold $\theta_d^x$. This threshold defines whether the search pattern is sufficiently similar to the template, so the pattern's central voxel may be a fiber voxel. For $d_{n}>\theta_d^x$ the corresponding voxels in $\mathbf O_x$ are set to 0. Note that all thresholds $\theta_d^x, \theta_d^y, \theta_d^z$ are not chosen arbitrarily, but derived adaptively from the input data. The detailed method of defining these thresholds will be described in a later section.

\item To all voxels being considered ($d_{n}<\theta_d^x$) a local minimum filter is applied, labeling only local best matching voxels as belonging to collagen phase by assigning the value 1, while all others are set to 0. After having finished this step the first x-slice (yz-plane) of the input data set is completely binarized and stored in the temporary binary output data set $\mathbf{O_x}$.

\item All previous steps are repeated for all other x-slices resulting the first binarized data set $\mathbf{O_x}$.

\item Some fibers with angles less than 45 degrees to any yz-plane (x-slice) may be not well detected since their cross-sections, appearing as roundish spots of finite size, are distorted and hence are not sufficiently similar to the template. Therefore the complete detection process is repeated in perpendicular xz- and xy-planes (y- and z-slices) of the sample volume. Having all y- and z-slices binarized two more binary data sets $\mathbf{O_y}$ and $\mathbf{O_z}$ are obtained.

\item Finally the three temporary reconstruction results are combined to one binary data set using logical OR:
\begin{equation}
O_{ijk} := O_{ijk}^x \vee O_{ijk}^y \vee O_{ijk}^z \ \ \ \ \forall \ i,j,k
\end{equation}

\item As a post-processing step all isolated voxels (i.e. solid phase voxels with all 26 neighbors belonging to fluid phase) are converted to fluid phase.

\end{enumerate}

\paragraph{Define optimal template sizes}

The right size $H\times W$ of the template $\mathbf{T} \in \mathds R^{H\times W}$ is crucial. On the one hand the template must not be too small, because otherwise the pattern is not completely covered. On the other hand templates should not be too large to limit computation time. A good indication is given by the algebraic sign of template's matrix entries. Since all entries are reduced by subtracting the mean gray scale $\mu$, positive entries correspond to gray scales brighter than and vice versa negative entries to gray scales darker than the average. Taking into account that the patterns to be detected are resulting from convolution of bright fibers with the point spread function it is reasonable to consider positive matrix entries as belonging to the pattern's foreground, while negative entries represent the surrounding background.
Hence, if the template is sized in a kind that all outer entries are negative and at the same time all inner entries are positive it is warranted that the pattern is completely covered by the template. Furthermore using both, positive and negative matrix entries, implies better use of the complete domain of definition and contrast enhancement since matching distances $d_n$ only will be minimal if the template is perfectly centered at the patterns being investigated.

Initially height $H$ and width $W$ are set to arbitrary, small values, e.g. $H=15$ and $W=9$. Then the templates are iteratively resized and recalculated until the optimal size is found. Since computing the template takes only a few seconds the complete runtime is not affected significantly by these iterations.

\paragraph{Define optimal matching thresholds}

The optimal thresholds $\theta_d^x$, $\theta_d^y$ and $\theta_d^z$ are also defined iteratively. It is clear that it depends on the thresholds how many voxels are labeled as fiber voxels. Because both templates and search matrices are normalized the maximum range of matching distances $d_n$ is $\left[0,2 \right]$ and consequently the optimal thresholds are in the same range as well.
\newline
If matrices are treated as vectors in high-dimensional feature space, three special cases can be distinguished:
\begin{description}
\item[Identity]
\noindent\hspace*{20mm} $d_n=0$ \noindent\hspace*{10mm} $\Rightarrow$ \noindent\hspace*{10mm} $\mathbf{S}_n \ \equiv \ \ \mathbf{T}$

\item[Orthogonality]
\noindent\hspace*{9.5mm} $d_n=\sqrt{2}$ \noindent\hspace*{7mm} $\Rightarrow$ \noindent\hspace*{10mm} $\mathbf{S}_n\ \ \bot \ \ \ \mathbf{T}$

\item[Inversion]
\noindent\hspace*{18mm} $d_n=2$ \noindent\hspace*{10mm} $\Rightarrow$ \noindent\hspace*{10mm} $\mathbf{S}_n \ =-\mathbf{T}$
\end{description}
Where orthogonality means a maximum dissimilarity between template and search matrix. Inversion implies identical absolute values of corresponding matrix entries with inverted algebraic signs.
Hence $\left[\sqrt{2},2 \right]$ is no expedient range for the thresholds which rather must be within $\left[0,\sqrt{2} \right]$.

Since reconstructed fibers should be skeletonized, the most probably number of direct fiber voxel neighbors $E_{mode}$ in a $3^3$-neighborhood of a central fiber voxel turns out to be a good criterion. Extensive evaluations of simulated line networks showed $E_{mode}=3$ in case of perfect skeletonization.

Obviously there is not only a single value for thresholds that achieves $E_{mode}=3$, but rather a range, where the optimal thresholds would be the top of this range, because this causes a maximum number of detected fibers while simultaneously the constraint $E_{mode}=3$ is fulfilled.
\newline
The threshold for each reconstruction direction x, y and z is defined in two subsequent steps. Firstly a threshold that fulfills $E_{mode}=4$ is found using a binary search algorithm. And secondly the found threshold is reduced step by step until it fulfills $E_{mode}=3$. 

\subparagraph{Binary Search} 

Binary search is an efficient standard algorithm for searching a specified value by halving the number of items to check with each iteration \cite{cor09, aro09}.
Initially the threshold is set to the middle of the range to be searched $\left[0,\sqrt{2} \right]$
\begin{equation}
\theta_d^{(0)}:=\frac{0+\sqrt{2}}{2}=\frac{1}{\sqrt{2}}
\end{equation}
and the bounds are defined
\begin{align}
\theta_d^{max(0)} & := \sqrt{2} \\
\theta_d^{min(0)} & := 0
\end{align}
\newline
Then the following steps are repeated until the stop criterion is fulfilled:
\begin{enumerate}
\item The fiber detection algorithm is performed with recent threshold $\theta_d^{(n)}$. To limit computation time not the complete data set (containing $512 \times 512 \times 597$ voxels) is used but a smaller sub set containing only $150 \times 150 \times 150$ voxels.

\item The most probable number of fiber voxel neighbors $E_{mode}$ is evaluated.

\item The threshold and the bounds of the searching range are updated: \\
If $E_{mode}<4$ then:
\begin{align}
\theta_d^{max(n+1)} & := \theta_d^{max(n)} \\
\theta_d^{min(n+1)} & := \theta_d^{(n)} \\
\theta_d^{(n+1)} & := (\theta_d^{max(n+1)}+\theta_d^{min(n+1)})/2
\end{align}
If $E_{mode}>4$ then:
\begin{align}
\theta_d^{max(n+1)} & := \theta_d^{(n)} \\
\theta_d^{min(n+1)} & := \theta_d^{min(n)} \\
\theta_d^{(n+1)} & := (\theta_d^{max(n+1)}+\theta_d^{min(n+1)})/2
\end{align}
If $E_{mode}=4$ then binary search is stopped.
\end{enumerate}

\subparagraph{Reduction of threshold}

The threshold is now iteratively reduced until $E_{mode}=3$:
\begin{enumerate}
\item The fiber detection algorithm is performed using the recent threshold $\theta_d^{(n)}$. Again due to limit computation time not the complete data set is used but a sub set. However containing now more voxels (i.e. $250 \times 250 \times 250$) than previously used for binary search to increase accuracy.

\item The most probably number of fiber voxel neighbors $E_{mode}$ is evaluated.

\item If $E_{mode}=4$ then the recent threshold is reduced by subtracting a small $\varepsilon$. \\
If $E_{mode}=3$ then the optimal threshold is found and the iteration loop is stopped. \\
Note that the smaller $\varepsilon$ is chosen the more exact the best threshold is found on the one hand but on the other hand the more iteration steps are required. As a good compromise to achieve both high accuracy and runtime limitation we found $\varepsilon=0.01$.
\end{enumerate}

\paragraph{Distribution of nearest obstacle distances}

The geometric properties of line networks, such as the pore sizes, are good indicators to estimate the similarity between different networks. The pore sizes of a network can be quantified in different ways, for instance by placing within each pore a sphere of the maximum possible size and then analyzing the size distribution of these spheres \cite{Mic08}. In this paper, we choose another, yet equivalent approach: We compute the distribution $p(r_{\rm{no}})$ of nearest obstacle distances in the binarized network. This is done by selecting a set of random points within the stack, computing the distance from each test point to its closest solid state obstacle (i.e. fiber segment) and then finding the distribution of these distances \cite{met11, ME-Paper}.

\paragraph{Quality measures}

To evaluate the validity of our algorithm we defined two quality measures based on Pearson product-moment correlation coefficient using the surrogate data sets. The \emph{sample correlation coefficient of local averaged voxel arrays} $r(\overline{S},\overline{R}) \in [-1,1]$ and the \emph{sample correlation coefficient of the distributions of nearest obstacle distances} $r(p_S,p_R) \in [-1,1]$. A value of $r(\overline{S},\overline{R})=1$ indicates a complete linear dependence between the two data sets and hence implies a perfect reconstruction of the network, while $r(\overline{S},\overline{R})=0$ corresponds to linear independence, i.e. both networks are entirely different. In a similar manner, $r(p_S,p_R)=1$ indicates that both distributions are identic. The reconstructed fibers, derived from grayscale surrogate data sets, are not exactly straight but rather smoothly fluctuating. That means that for a given solid phase voxel in binary surrogate data set the position of the corresponding solid phase voxel in the reconstructed data set may differ. However, the range of difference does not exceed one voxel size in each direction. Taking into account that such small fluctuations do not affect the network's global properties, we do not compare the binary data sets (i.e. binary surrogate and binary reconstruction) itselves, but local averaged voxel arrays $\overline{S}$ and $\overline{R}$. Therefore both binary data sets to be compared are converted by setting each voxel to the average value of itself and its 26 direct neighbors. The resulting arrays provide some significant advantages: The information of local solid voxel density is preserved since larger average values are corresponding to a larger number of solid phase voxels within a $3^3$-neighborhood. Furthermore, independency against small differences from exact positions is achieved and the global solid voxel distribution, i.e. the network morphology, is also preserved. After having converted the data sets to be compared, the sample correlation coefficient $r(\overline{S},\overline{R})$ of voxel values is calculated. To compare the distributions of nearest obstacle distances $p_S(r_{\rm{no}})$ and $p_R(r_{\rm{no}})$ in binary surrogate and reconstruction result, firstly both distributions are evaluated and secondly the empirical correlation coefficient $r(p_S,p_R)$ is calculated.

\section*{Results}

\begin{figure} [!htb]
\begin{center}
\includegraphics[width=0.5\linewidth]{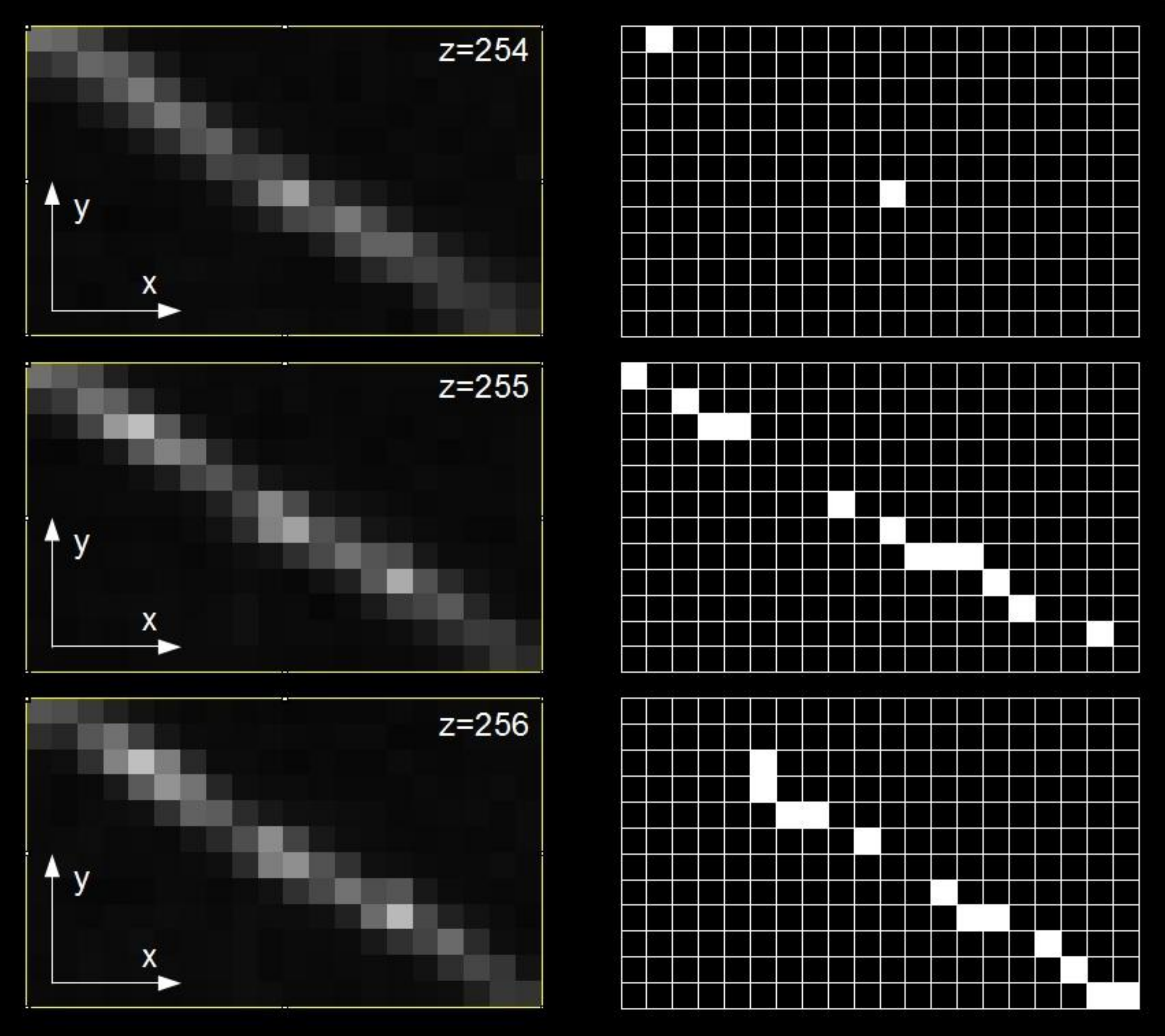}
\end{center}
\caption{\textbf{Reconstruction result.} Three subsequent original microscope images (left) and the corresponding binarizations (right) generated by the template matching algorithm.}
\label{fig:recresults}
\end{figure}

In the following, we show the results of binarizing grayscale image stacks with the template matching algorithm, i.e. the insensitivity to variations in the input data quality and the correct reconstruction of known networks. A typical reconstruction result of original microscope image stacks can be seen in Fig.\,\ref{fig:recresults}. 

\subsection*{Insensitivity to variations in the input data quality}

\begin{figure} [!htb]
\begin{center}
\includegraphics[width=0.8\linewidth]{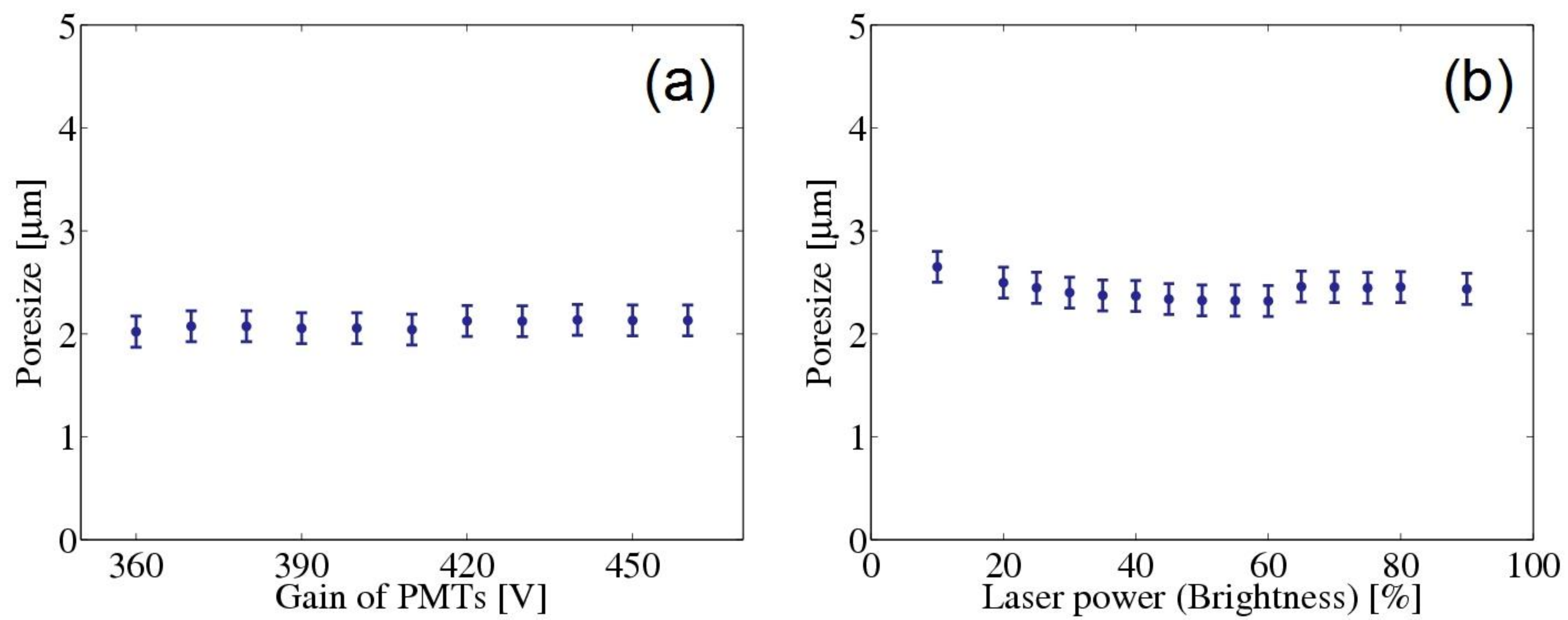}
\end{center}
\caption{\textbf{Insensitivity of the algorithm to variations in the input data quality.} The algorithm produces stable results in a wide range of photomultiplier gain and laser outlet power. Hence it is insensitive to variations in the input data quality. (a) Evaluated pore size as a function of photomultiplier gain (signal-to-noise-ratio). (b) Evaluated pore size as a function of laser outlet power (image brightness). The data in (a) and (b) correspond to two collagen gels that have been fabricated under identical conditions. The slight differences in the observed pore sizes reflect natural sample-to-sample fluctuations.}
\label{fig:INDIPEND}
\end{figure}

We first showed that our binarization algorithm was widely independent from the setup parameters of the imaging process, such as the laser outlet power of the confocal microscope that defines the global picture brightness and the gain of the photomultiplier tubes (PMTs) which mostly determines the signal-to-noise-ratio (SNR) of the images.
Therefore, both parameters were changed systematically in the available range. The laser outlet power was shifted from 10\,\% to 90\,\% of the laser's produced beam, the gain being constantly fixed to an apropriate value for 50\,\% laser outlet power. Furthermore, the gain was varied about 50\,V around the fitting value for a fixed laser power of 50\,\%.
The results of a sequence of changed parameters can be seen in Fig.\,\ref{fig:INDIPEND}.

\subsection*{Correct reconstruction of known networks}

We calculated $r(\overline{S},\overline{R})$ to evaluate the similarity between the surrogate and the reconstructed networks (compare Methods Section). 

To highlight the performance of the template matching algorithm (PM), we also calculated $r(\overline{S},\overline{R})$ for a simple threshold binarization algorithm (TH). We used a threshold $\theta=\mu+2\sigma$, with mean grayscale $\mu$ and standard deviation of grayscales $\sigma$, since extensive tests have shown, that this rough rule of thumb provides a quite fair value for the threshold.

Typical results are:
\begin{quote}
	$r_{PM}(\overline{S},\overline{R})=0.84$ \ \ and \ \ $r_{TH}(\overline{S},\overline{R})=0.46$
\end{quote}

\begin{figure} [!htb]
\begin{center}
\includegraphics[width=0.4\linewidth]{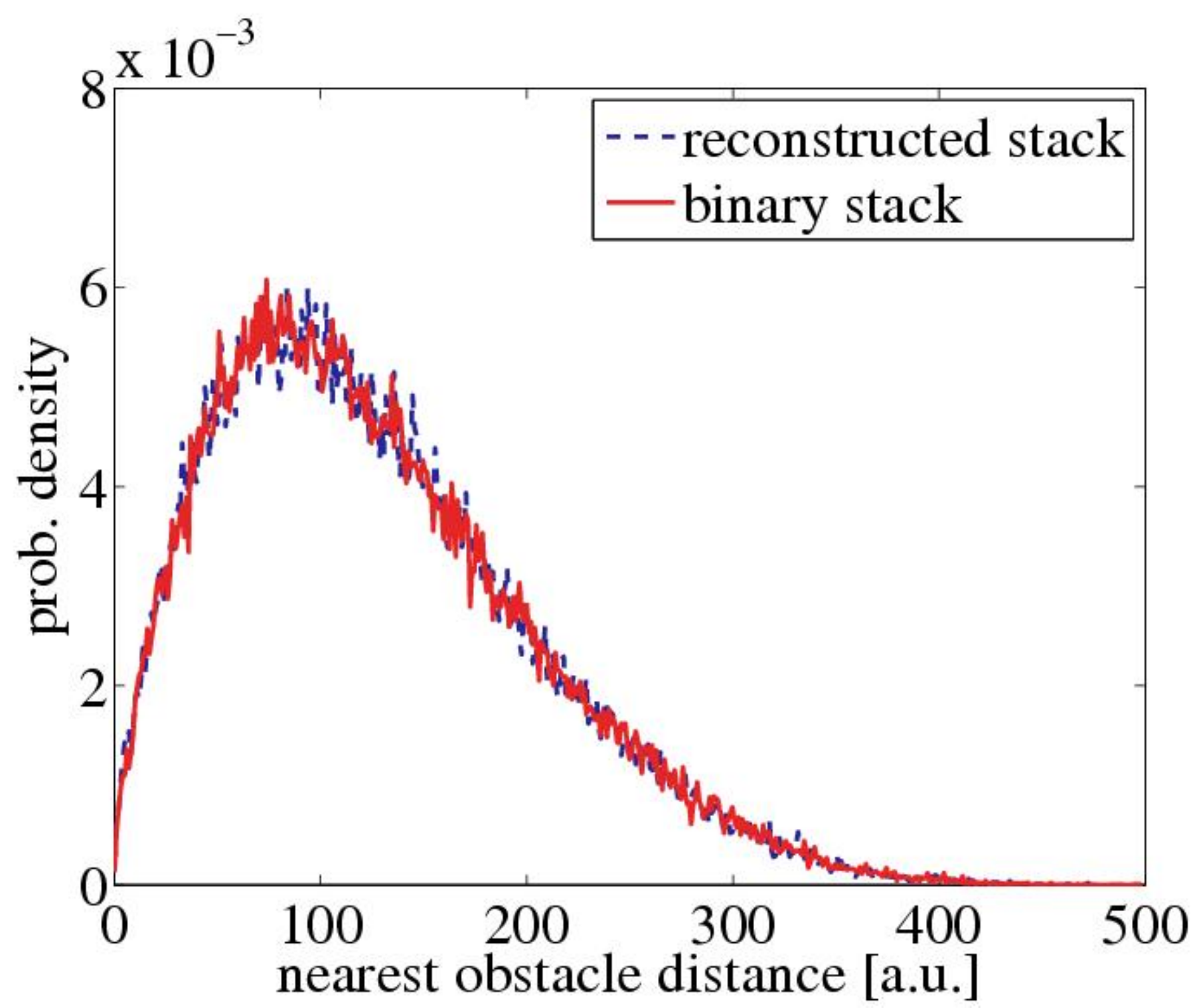}
\end{center}
\caption{\textbf{Distributions of nearest obstacle distances $p_S(r_{\rm{no}})$ and $p_B(r_{\rm{no}})$ in binary surrogate data set and reconstruction result.} Both distributions are, disregarding statistical fluctuations, almost identic.}
\label{fig:quality}
\end{figure}

The algorithm's correct reconstruction was proved by using surrogate data sets. We compared the distributions $p_S(r_{\rm{no}})$ and $p_B(r_{\rm{no}})$ of nearest obstacle distances for about 100 surrogate and reconstructed networks and calculated the correlation coefficient. As can be seen in the example of Fig.\,\ref{fig:quality}, the distributions are almost identical with a sample correlation coefficient of $0.93$.

\subsection*{Summary and Outlook}

In this paper we have presented a fast, robust and objectively tested method to reconstruct disordered fiber networks from confocal image stacks. In the original stacks, visible fiber segments appear as "cylindrical clouds" with a bright core, surrounded by a broad "halo" with slowly decaying gray level. After reconstruction, the fiber segments are represented by contiguous voxels with value 1, while all background voxels are assigned the value 0. These reconstructed traces, due to the remaining voxelization, "wiggle" around the smooth space curve of the fiber's medial axis. This level of reconstruction is sufficient for many purposes, such as the statistical evaluation of the distribution of nearest obstacle distances in the fiber network.

However, other statistical investigations, for example evaluating the distribution of curvatures along the fibers, would benefit from a parametrized description of each visible fiber segment $s$ in terms of a space curve $\vec{R}_s(t)$. This could be achieved by a suitable post-processing of the present, voxelized representation. Alternatively, our template matching algorithm could be extended to sub-voxel accuracy. In this case the 2D position of the fiber center would be treated as a continuous variable within each of the cross sectional planes. The local mismatch minimum can then be found, with arbitrary spatial resolution, using standard continuum optimization techniques. Once the fiber centers are determined in each cross sectional plane, they can be connected by straight line segments or spline-interpolated to obtain the space curves $\vec{R}_s(t)$. 

\section*{Acknowledgments}

We are grateful for financial support by the German Research Foundation (DFG).

\clearpage
\newpage

\section*{References}
\renewcommand\refname{}
\bibliography{references}





\end{document}